\documentstyle[aps,prl,epsf]{revtex}

\def\r#1{\ignorespaces $^{#1}$}

\begin{document}

\title{\vspace*{-1cm} 
 The $\mu\tau$ and $e\tau$ Decays of Top Quark Pairs Produced in 
 $p\overline{p}$ Collisions at $\sqrt{s} = 1.8$ TeV}

\author{
\font\eightit=cmti8
\hfilneg
\noindent
F.~Abe,\r {17} H.~Akimoto,\r {36}
A.~Akopian,\r {31} M.~G.~Albrow,\r 7 S.~R.~Amendolia,\r {27} 
D.~Amidei,\r {20} J.~Antos,\r {33} S.~Aota,\r {36}
G.~Apollinari,\r {31} T.~Asakawa,\r {36}
% ...
\vskip .026in
\begin{center}
(CDF Collaboration)
\end{center}
\vskip .026in
\begin{center}
\r 1  {\eightit Argonne National Laboratory, Argonne, Illinois 60439} \\
\r 2  {\eightit Istituto Nazionale di Fisica Nucleare, University of Bologna,
I-40127 Bologna, Italy} \\
\r 3  {\eightit Brandeis University, Waltham, Massachusetts 02264} \\
\r 4  {\eightit University of California at Los Angeles, Los 
Angeles, California  90024} \\  
\r 5  {\eightit University of Chicago, Chicago, Illinois 60638} \\
\r 6  {\eightit Duke University, Durham, North Carolina  28708} \\
\r 7  {\eightit Fermi National Accelerator Laboratory, Batavia, Illinois 
60510} \\
\r 8  {\eightit University of Florida, Gainesville, FL  33611} \\
\r 9  {\eightit Laboratori Nazionali di Frascati, Istituto Nazionale di Fisica
               Nucleare, I-00044 Frascati, Italy} \\
\r {10} {\eightit University of Geneva, CH-1211 Geneva 4, Switzerland} \\
\r {11} {\eightit Harvard University, Cambridge, Massachusetts 02138} \\
\r {12} {\eightit Hiroshima University, Higashi-Hiroshima 724, Japan} \\
\r {13} {\eightit University of Illinois, Urbana, Illinois 61801} \\
\r {14} {\eightit Institute of Particle Physics, McGill University, Montreal 
H3A 2T8, and University of Toronto,\\ Toronto M5S 1A7, Canada} \\
\r {15} {\eightit The Johns Hopkins University, Baltimore, Maryland 21218} \\
\r {16} {\eightit Universit\"{a}t Karlsruhe, 76128 Karlsruhe, Germany} \\
\r {17} {\eightit National Laboratory for High Energy Physics (KEK), Tsukuba, 
Ibaraki 315, Japan} \\
\r {18} {\eightit Ernest Orlando Lawrence Berkeley National Laboratory, 
Berkeley, California 94720} \\
\r {19} {\eightit Massachusetts Institute of Technology, Cambridge,
Massachusetts  02139} \\   
\r {20} {\eightit University of Michigan, Ann Arbor, Michigan 48109} \\
\r {21} {\eightit Michigan State University, East Lansing, Michigan  48824} \\
\r {22} {\eightit University of New Mexico, Albuquerque, New Mexico 87132} \\
\r {23} {\eightit The Ohio State University, Columbus, OH 43320} \\
\r {24} {\eightit Osaka City University, Osaka 588, Japan} \\
\r {25} {\eightit Universita di Padova, Istituto Nazionale di Fisica 
          Nucleare, Sezione di Padova, I-36132 Padova, Italy} \\
\r {26} {\eightit University of Pennsylvania, Philadelphia, 
        Pennsylvania 19104} \\   
\r {27} {\eightit Istituto Nazionale di Fisica Nucleare, University and Scuola
               Normale Superiore of Pisa, I-56100 Pisa, Italy} \\
\r {28} {\eightit University of Pittsburgh, Pittsburgh, Pennsylvania 15270} \\
\r {29} {\eightit Purdue University, West Lafayette, Indiana 47907} \\
\r {30} {\eightit University of Rochester, Rochester, New York 14628} \\
\r {31} {\eightit Rockefeller University, New York, New York 10021} \\
\r {32} {\eightit Rutgers University, Piscataway, New Jersey 08854} \\
\r {33} {\eightit Academia Sinica, Taipei, Taiwan 11530, Republic of China} \\
\r {34} {\eightit Texas A\&M University, College Station, Texas 77843} \\
\r {35} {\eightit Texas Tech University, Lubbock, Texas 79409} \\
\r {36} {\eightit University of Tsukuba, Tsukuba, Ibaraki 315, Japan} \\
\r {37} {\eightit Tufts University, Medford, Massachusetts 02155} \\
\r {38} {\eightit University of Wisconsin, Madison, Wisconsin 53806} \\
\r {39} {\eightit Yale University, New Haven, Connecticut 06511} \\
\end{center}}

\maketitle

\begin{abstract}
We present evidence for dilepton events from $t\overline{t}$ production with
one  electron or muon and one hadronically decaying $\tau$ lepton from the
decay $t\overline{t}\ \rightarrow\ (\ell\nu_{\ell}) (\tau\nu_\tau)\
b\overline{b}$,  ($\ell=e, \mu$), using the Collider Detector at Fermilab
(CDF). In a 109 pb$^{-1}$ data sample  of $p\overline{p}$ collisions at 
$\sqrt{s} = 1.8$ TeV  we expect $\sim$ 1 signal event and a total background of
$\sim$ 2 events; we observe 4 candidate events (2 $e\tau$ and 2 $\mu\tau$).
Three of these events have jets identified  as $b$ candidates, compared to an
estimated background of $0.28\pm0.02$ events. 
\end{abstract} 

\begin{center}
(submitted to PRL)
\end{center}

\narrowtext
\twocolumn 
The Collider Detector at Fermilab (CDF) Collaboration~\cite{top_prd,top_prl} 
and the D0 Collaboration~\cite{top_D0} recently established the existence of 
the top quark through searches for $t\overline{t}$ production with the
subsequent decay $t\overline{t} \rightarrow W^+b W^-\overline{b}$. The decay
modes of the two $W$ bosons determine the observed event signature. Both
experiments observed top quarks based on the ``dilepton'' channels in which
both $W$ bosons decay into $e\nu_e$ or $\mu\nu_\mu$, and the ``lepton + jets''
channel where one $W$ boson  decays into $e\nu_e$ or $\mu\nu_\mu$ and the other
into quarks.

Here we present first evidence for top quark decays in the ``tau dilepton''
channel, where one $W$ decays into $e\nu_e$ or $\mu\nu_\mu$ and the other into
the third-generation leptons, $\tau$ and $\nu_\tau$. Consequently, the total 
decay chain is:
%
\begin{center}
$t\overline{t}\rightarrow W^+ W^- b\overline{b}
\rightarrow (\ell\nu_{\ell}) (\tau\nu_{\tau})\ b\overline{b}$ \vspace*{1mm}\\
\hspace*{4.15cm}$^{|} \vspace*{-4mm}$ \newline
\hspace*{5.2cm}$\rightarrow hadrons + \nu_{\tau},$ 
\end{center}
where $\ell$ stands for $e$ or $\mu$. This channel is of particular interest
because the existence of a charged Higgs boson $H^\pm$ with $m_{H^\pm}<m_{top}$ 
could give rise to anomalous $\tau$ lepton production through the decay 
chain $t\rightarrow H^+ b\rightarrow \tau^+\nu_{\tau} b$,
which could be directly observable in this channel~\cite{higgs}.

In the Standard Model the top branching ratio (BR) to $Wb$ is  essentially
100\% and the approximate BR of $W$ to each of $e\nu_e$, $\mu\nu_\mu$, and 
$\tau\nu_\tau$ is $1/9$, and to $q\overline{q}^\prime$ is $6/9$. Consequently,
the  total BR for $t\overline{t}$ into $e\tau$ and $\mu\tau$ events is 4/81,
the same as for $ee$, $\mu\mu$, and $e\mu$ combined. In principle, the  number
of dilepton events could be doubled by including $\tau$'s.  However, the 64\%
BR\cite{pdg} for $\tau$ decays into hadrons (50\% one-prong and 14\%
three-prong decays), decreased kinematic acceptance due to the undetected
$\nu_\tau$, and a $\tau$ selection that is less efficient than the $e$ or $\mu$
selection, result in a total tau dilepton acceptance about five times smaller
than that for $ee,\ \mu\mu$, and $e\mu$ events.

We report here on a search based on a data sample containing $109\pm7$
pb$^{-1}$  collected with CDF during the Fermilab 1992-93 and 1994-95 Collider
runs. A detailed description of the detector can be found elsewhere~\cite{cdf}.
The components of the detector most relevant to this search are a four-layer
silicon vertex detector~\cite{svx_prime}, located immediately outside the beam
pipe, providing precise track reconstruction used to identify secondary
vertices from $b$ and $c$ quark decays, a central drift chamber immersed in a
1.4 T solenoidal magnetic field for tracking charged particles in the
pseudorapidity~\cite{eta} range $|\eta|<1.1$, electromagnetic and hadronic
calorimeters covering the range $|\eta|<4.2$ and arranged in a projective tower
geometry for identifying electrons and jets, strip chambers embedded in the
electromagnetic calorimeter at a depth of approximately shower maximum for
detailed shower sampling, and drift chambers outside the calorimeters in the
region $|\eta|<1.0$ for muon identification. Calorimeters also measure the
missing transverse energy $E\!\!\!\!/_T$\cite{eta}, which can indicate the
presence of undetected energetic neutrinos. A three-level trigger selects
inclusive electron and muon events used in this analysis.

The data sample used in this analysis comprises \mbox{high-$p_T$} inclusive
lepton events that contain an electron with $E_T> 20$ GeV or a muon with
$p_T>20$ GeV/$c$ in the central region ($|\eta|<1.0$). The selection criteria
for the primary $e$ or $\mu$ are identical to those applied 
in Ref.~\cite{top_prl}.  

The identification of hadronically decaying $\tau$'s  is difficult due to
the misidentification of the much more numerous quark or gluon jets as $\tau$'s.
We use two complementary techniques for identifying $\tau$'s, one
``track--based'' and the other ``calorimeter--based''.

The track--based selection~\cite{michele} accepts only one-prong $\tau$ decays.
Events with an $e$ or a $\mu$ must have an additional  high-$p_T$ ($p_T > 15$
GeV/$c$), central ($|\eta|<1.0$), isolated track. The tracking isolation
$I_{track}$ is defined as $\Sigma p_T$ of all tracks in a cone of $\Delta
R=\sqrt{(\Delta\phi)^2+(\Delta\eta)^2}=0.4$ in $(\eta,\phi)$ space around the
high-$p_T$ track. A cut of \mbox{$I_{track} < 1$ GeV/$c$} discriminates between
the $\tau$ signal and QCD jets.  Requiring $E/p>0.5$ ensures consistency
between the energy measured in the calorimeter and the track momentum.
Electrons are removed by rejecting clusters in which a large fraction of the
total  energy is deposited in the electromagnetic calorimeter. Tracks
associated with an energy deposition consistent with that of a minimum ionizing
particle are rejected as $\mu$ candidates. These cuts provide sufficient
background rejection for one-prong decays, but not for three-prong decays.

The calorimeter--based selection~\cite{marcus} increases the acceptance by 
using both one-prong and three-prong $\tau$ decays.
The selection criteria are:
(i) The number of tracks with $p_T>1$ GeV/$c$ in a $10^\circ$ cone around the 
calorimeter cluster center, which defines the track multiplicity associated 
with the cluster, must be either one or three.
(ii) The track isolation $I_{track}$ is defined as $\Sigma p_T$ of all tracks in
a cone of $\Delta R=0.4$ around the cluster center, excluding those that 
define the track multiplicity. We require \mbox{$I_{track}<1$ GeV/$c$}.
(iii) About 73\% (41\%) of all one(three)-prong decays are expected to be 
associated with at least one $\pi^0$\cite{pdg} which is identifiable 
in the strip chambers by searching for
clusters from the decay $\pi^0 \rightarrow \gamma\gamma$. 
The $p_T$  of the $\tau$ is then defined as the scalar sum of the $p_T$  of the 
tracks in the $10^\circ$ cone plus the $E_T$ of any identified $\pi^0$'s as 
measured in the electromagnetic calorimeter.
We require $p_T > 15$ GeV/$c$ and $|\eta |<1.2$. 
(iv) We require $0.5 < E/p < 2.0(1.5)$ for one(three)-prongs.
(v) The width $\sigma_{cl}$ of a calorimeter cluster in $(\eta,\phi)$ space is
defined as the second moment of the $E_T$ distribution among the towers in a 
cluster. Low--multiplicity $\tau$ clusters are narrower than clusters from QCD 
jets: we require $\sigma_{cl} < 0.11(0.13) - 0.025(0.034)\times 
E_T[\mbox{GeV}]/100$ for one(three)-prongs.
(vi) Tau decays rarely involve more than 2 $\pi^0$'s, so fewer than 
3 $\pi^0$ candidates must be found.  
(vii) The invariant mass reconstructed from 
tracks and $\pi^0$'s is required to be less than  1.8 GeV/$c^2$. 
(viii) Clusters consistent with being an $e$ or $\mu$ are removed.

A Monte Carlo simulation ($m_{top} = 175\ \mbox{GeV}/c^2$) of $t\overline{t}$
production provides an estimate of the $\tau$ identification efficiencies and
acceptances for tau dilepton events. We use the {\sc pythia}~\cite{pythia}
Monte Carlo to generate $t\overline{t}$ events, the {\sc tauola}
package~\cite{tauola}, which  correctly treats the $\tau$ polarization, to
decay the tau lepton, and a detector simulation. We expect 29\% of hadronic
one-prong $\tau$ decays to produce tracks with \mbox{$p_T > 15$ GeV/$c$} and
$|\eta |<1.0$. The track--based $\tau$ selection identifies
(59$\pm$4(stat)$\pm$3(syst))\% of these. The calorimeter--based selection
identifies \mbox{(57$\pm$2(stat)$\pm$3(syst))\%} of the 45\% of all hadronic
$\tau$ decays with \mbox{$p_T > 15$ GeV/$c$}, as defined in the previous
paragraph, and $|\eta|<1.2$. The uncertainty in the number of tracks due to the
underlying event and  overlapping minimum bias events makes the largest
contribution to the overall systematic uncertainty.

\begin{figure}
\epsfysize=8.5cm 
\vspace*{-1mm} 
\hspace*{10mm}
\epsfbox{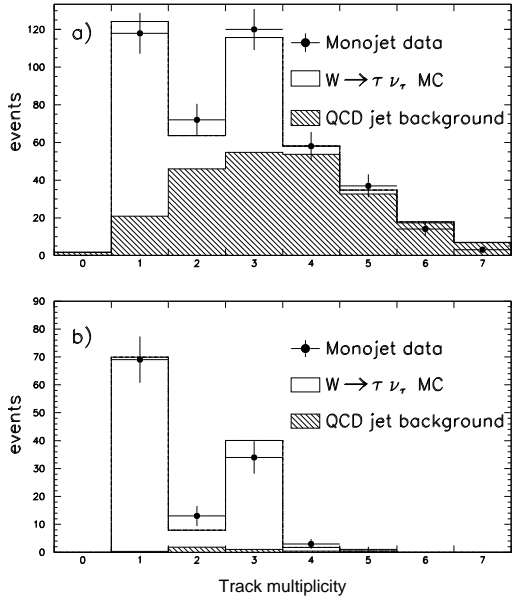}
\vspace*{1mm} 
\caption{Track multiplicity in the monojet data sample. a) No $\tau$ ID cuts 
applied. b) After applying all $\tau$ ID cuts except track multiplicity.}
\label{monojet} 
\end{figure}

The efficiency calculation is checked using a data sample enriched in 
$W\hspace*{-1mm}\rightarrow\hspace*{-1mm}\tau\nu_\tau$ decays. Typically, a 
$W\hspace*{-1mm}\rightarrow\hspace*{-1mm}\tau\nu_\tau
\hspace*{-1mm}\rightarrow\hspace*{-1mm} hadrons + \nu_\tau 
\overline{\nu}_\tau$ decay has one jet from the $\tau$, and $E\!\!\!\!/_T$ due
to the neutrinos. A monojet sample is selected by requiring one central jet
with \mbox{$15 < E_T <40$ GeV}, no other jet with  \mbox{$E_T$ $>7$ GeV} in
\mbox{$|\eta|<4.0$}, and \mbox{$20<$ $E\!\!\!\!/_T$ $<40$ GeV}.
Figure~\ref{monojet}a shows the track multiplicity in this sample and in a
background sample of QCD jets. The latter is normalized to the monojet sample
using the bins with $\geq 4$ tracks where there is a very small contribution
from $W\hspace*{-1mm}\rightarrow\hspace*{-1mm}\tau\nu_\tau$ events. The data
show a clear excess in the one-prong and three-prong bins, as expected for a
sample with significant $\tau$ fraction. The
$W\hspace*{-1mm}\rightarrow\hspace*{-1mm}\tau\nu_\tau$ content is estimated to
be (45$\pm$5(stat))\% by subtracting the QCD contribution.
Figure~\ref{monojet}b shows the track multiplicity after applying all  cuts
from the calorimeter--based $\tau$ selection (except cut i). The background in
all bins is greatly  
\noindent reduced and the data agree well with the expectation from 
a $W\hspace*{-1mm}\rightarrow\hspace*{-1mm}\tau\nu_\tau$ Monte
Carlo~\cite{pythia}. The efficiency of all calorimeter--based $\tau$
identification cuts is measured to be  (55$\pm$6(stat))\%, consistent with the
$W\hspace*{-1mm}\rightarrow\hspace*{-1mm}\tau\nu_\tau$ Monte Carlo prediction
of (56$\pm$1(stat))\%.  The same check performed on the track--based $\tau$
selection gives similar results.

Top events and background have different topologies. Dilepton events from
$t\overline{t}$ decays are expected to contain 2 jets  from $b$ decays. We
therefore select events with $\geq 2\ \mbox{jets  with}\  E_T > 10$ GeV and
\mbox{$|\eta| < 2.0$}~\cite{top_prl}.  The $E\!\!\!\!/_T$ is corrected for
muons and jets as in the dilepton  analysis\cite{top_prd,top_prl}. As top
events are expected to have significant amount of $E\!\!\!\!/_T$ due to
undetected neutrinos, a cut is applied on the $E\!\!\!\!/_T$  significance,
defined for $e\tau$ events as $S_{E\!\!\!\!/_T}\  \equiv
\frac{E\!\!\!\!/_T} {\sqrt{\Sigma E_T}},$ and as  
$S_{E\!\!\!\!/_T}\ \equiv \frac{E\!\!\!\!/_T} {\sqrt{\Sigma E_T
+p_T^\mu}}$ for $\mu\tau$ events. Here $\Sigma E_T$ is the  scalar sum of the
transverse  energies measured in the calorimeter towers. We  require
$S_{E\!\!\!\!/_T}\ >3\ \mbox{GeV}^{1/2}$.  Due to large  $m_{top}$,
$t\overline{t}$ events exhibit large total transverse energy,  $H_T$~\cite{ht}.
We require   $H_T \equiv E_T^{e} (p_T^{\mu}) + p_T^{\tau} +
E\!\!\!\!/_T\hspace*{3mm} + (\sum_{jets} E_T)>180$ GeV. Finally, the
leptons must have opposite charge.

The product of all BR's, geometric and kinematic acceptance,  efficiencies for
trigger, lepton identification, and cuts on the event  topology yields a total
acceptance \mbox{$A_{tot} = (0.085\pm$0.010(stat)$\pm$0.012(syst))\%}  for the
track--based selection. Using the calorimeter-based selection we find {$A_{tot}
=  (0.134\pm$0.013(stat)$\pm$0.019(syst))\%}. The systematic uncertainty on
$A_{tot}$ is dominated by uncertainties on identification efficiencies for the
$\tau$ (6\%) and the primary lepton (7\%),  the top mass (6\%), and the
hadronic energy scale of the calorimeter (5\%). Of the total one-prong events
selected, 19\% (38\%) are expected to be found  only by the
track(calorimeter)--based technique, and 43\% by both. Based on the
$t\overline{t}$ cross section as measured by CDF from other decay modes,  we
expect 0.7$\pm$0.2(stat)$\pm$0.1(syst) and 1.1$\pm$0.3(stat)$\pm$0.2(syst) 
events from $t\overline{t}$ production in the two selections, respectively.

Table~\ref{background_relax} lists the contributions from the various
background sources. The dominant background is due to  $Z/\gamma\rightarrow
\tau^+ \tau^- + jets$ events. If one $\tau$ decays leptonically  and the other
$\tau$ hadronically, this process can mimic the top signature. From Monte Carlo
simulations we  expect  a background of 0.89$\pm$0.28 (1.48$\pm$0.38) events
due to this process for the track(calorimeter)--based $\tau$  selection,  and
smaller backgrounds from $WW$ and $WZ$ production.

The ``fake $\tau$'' background is due to $W+\geq 3\ jets$ events with one jet 
misidentified as a $\tau$. We calculate the fake rates as a function of $E_T$
by applying the $\tau$ selection criteria to jets in QCD jet samples. Applying
the fake rates bin-by-bin to the $E_T$ spectrum of all jets that could be
misidentified as $\tau$'s in a $W+\geq 3\ jets$ sample gives the number of
fake events. We expect 0.25$\pm$0.02 fake one-prong $\tau$'s with the  
track--based $\tau$ selection, and 0.78$\pm$0.04 fake one- and three-prong
$\tau$'s  with the calorimeter--based selection. The total expected backgrounds
are 1.28$\pm$0.29 and 2.50$\pm$0.43 events, respectively. 

We check that our background calculations correctly predict the number of
events in a background-dominated sample by dropping the $H_T$ and the
$S_{E\!\!\!\!/_T}$ requirements and instead imposing a loose $E\!\!\!\!/_T$ cut
($E\!\!\!\!/_T$ $>15$ GeV). With these relaxed cuts we expect a total
background of $5.7\pm0.7$ ($9.4\pm0.8$) events, in addition to 1.3 (2.0) events
from $t\overline{t}$ decays, and observe 9 (11) events in the data using the
track(calorimeter)--based selection. The sum of calculated background and top
contribution agrees well with the observed number of events. 

\begin{figure}
\epsfxsize=7.5cm
\hspace*{4mm}
\epsfbox{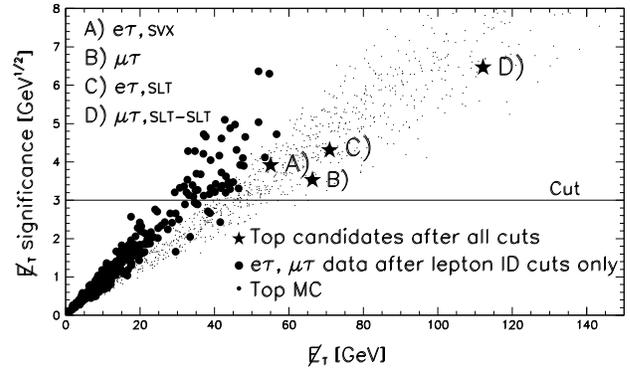}
\vspace*{-5.2cm}
\caption{The distribution of $S_{E\!\!\!\!/_T}$ vs $E\!\!\!\!/_T$ for events 
with a primary lepton and a tau candidate in the data. Three of the four final 
candidate events (stars) have $b$-tagged jets.}
\label{metsig_vs_met_data} 
\end{figure}

Figure~\ref{metsig_vs_met_data} shows $S_{E\!\!\!\!/_T}$ versus $E\!\!\!\!/_T$
for data events  with a primary lepton and a tau candidate that passes the
calorimeter--based selection cuts. After all cuts four candidate events are
identified, 2 $e\tau$ and 2 $\mu\tau$ events. There is in addition one
same-sign $\mu^+\tau^+$ event, consistent with the 0.78 expected background
events from fake $\tau$'s. The track--based $\tau$ selection finds the same
four events. 

We use the presence of a soft lepton from semileptonic $b$ decays (SLT) or of a
secondary vertex (SVX) in the silicon vertex detector to identify jets from $b$
quarks. Three of the four candidate events have $b$-tagged jets~\cite{top_prl}.
One event has an SLT-SLT double tag. We expect 0.16 (0.18) background events
with $\geq 1$ SVX (SLT) tag, for a total background including correlations of
0.28$\pm$0.02 events. The probability to observe $\geq 3$ background events is
0.3\% after \mbox{$b$-tagging}.  For top signal plus background we expect
0.64$\pm$0.12(stat) (0.37$\pm$0.06(stat)) events  with an SVX(SLT) tag and 
observe one (two) event(s). 

In conclusion, we have developed a method to use $\tau$ leptons in the analysis
of top decays. We observe 4 candidate events where we expect $\sim$ 1
$t\overline{t}$  event and $\sim 2$ background events. In three of the
events we identify  jets from $b$ quark decays, which supports the
$t\overline{t}$ hypothesis.  Using the numbers of  estimated background and
observed events in Table~\ref{background_relax} ($N_{jet}\geq 2$) and the
acceptances $A_{tot}$, we calculate a production cross section. We find
\mbox{$\sigma_{t\overline{t}} = 10.2^{+16.3}_{-10.2}$(stat)$\pm$1.6(syst) pb}
for the calorimeter--based selection and 
\mbox{$29.1^{+26.3}_{-18.4}$(stat)$\pm$4.7(syst) pb} for the track--based
selection, consistent with latest measured values given the large statistical
uncertainty.

We thank the Fermilab staff and the technical staffs of the participating
institutions for their vital contributions. 
\widetext
\begin{table*}[t]
\caption{The expected number of background and $t\overline{t}$ events and 
the observed events.} 
\label{background_relax} 
\begin{center}
\begin{tabular}{lcccc}
Selection                    & \multicolumn{2}{c}{Track--based} &  
                               \multicolumn{2}{c}{Calorimeter--based} \\
$N_{jet}\ (\geq 10\mbox{ GeV})$&       1        &      $\geq 2$ 
                             &        1        &      $\geq 2$  \\ \hline
$\tau$ fakes                 &   0.14$\pm$0.01 & 0.25$\pm$0.02  
                             &   0.47$\pm$0.03 & 0.78$\pm$0.04  \\
$Z/\gamma\rightarrow \tau^+ \tau^-$ 
                             &   0.22$\pm$0.12 & 0.89$\pm$0.28  
                             &   0.54$\pm$0.16 & 1.48$\pm$0.38  \\
$WW,\ WZ$                    &   0.14$\pm$0.06 & 0.14$\pm$0.08  
                             &   0.20$\pm$0.09 & 0.24$\pm$0.10  \\
Total Background             &   0.50$\pm$0.14 & {\bf 1.28$\pm$0.29}  
                             &   1.21$\pm$0.28 & {\bf 2.50$\pm$0.43}  \\ \hline
expected from $t\overline{t}$&   0.08$\pm$0.02 & 0.7$\pm$0.3  
                             &   0.13$\pm$0.03 & 1.1$\pm$0.4    \\ \hline
observed events ($b$-tagged events) &  1 (0) & 4 (3) 
                                    &  0 (0) & 4 (3)      \\ 
\end{tabular} 
\end{center}
\end{table*}
\narrowtext
\noindent This work was supported by the U.S.
Department of Energy and National Science Foundation; the Italian Istituto
Nazionale di Fisica Nucleare; the Ministry of Education, Science and Culture
of Japan; the Natural Sciences and Engineering Research Council of Canada; the
National Science Council of the Republic of China, and the A.P. Sloan
Foundation.                                 
\vspace{-5mm}

\begin{thebibliography}{99}
\bibitem{eta} In the CDF coordinate system, $\theta$ and $\phi$ are the 
polar and azimuthal angles, respectively, with 
respect to the proton beam direction. The pseudorapidity $\eta$ is defined as
$-\ln\tan(\theta/2)$. The transverse momentum of a particle is $P_T = P
\sin\theta$. The analogous quantity using energies, defined as $E_T = E
\sin\theta$, is called transverse energy.
The missing transverse energy $E\!\!\!\!/_T$ is defined as $-\sum E_T^i\ 
\hat{r}_i$, where $\hat{r}_i$ are the unit vectors in the transverse plane 
pointing to the energy depositions in the calorimeter.
\end{thebibliography}
\end{document}